# Electrically Tunable Multifunctional All-Dielectric Metasurfaces Integrated with Liquid Crystals in the Visible

Yueqiang Hu[1,3], Xiangnian Ou[1], Tibin Zeng[2], Jiajie Lai[1], Jian Zhang[1], Xin Li[1], Xuhao Luo[1], Ling Li[1], Fan Fan[2*], and Huigao Duan[1,3*]

[1] National Research Center for High-Efficiency Grinding, College of Mechanical and Vehicle Engineering, Hunan University, Changsha 410082, P.R. China

[2] Key Laboratory for Micro/Nano Optoelectronic Devices of Ministry of Education & Hunan Provincial Key Laboratory of Low-Dimensional Structural Physics and Devices, School of Physics and Electronics, Hunan University, Changsha 410082, China

[3] Advanced Manufacturing Laboratory of Micro-Nano Optical Devices, Shenzhen Research Institute, Hunan University, Shenzhen, 518000, China

*Corresponding authors. Email: duanhg@hnu.edu.cn, ffan@hnu.edu.cn

**Abstract:**

As two-dimensional metamaterials, metasurfaces open up new avenues for designing static planar optics. However, the dynamic modulation of metasurfaces in the optical band is required for practical applications. The existing dynamic devices rarely utilized the polarization manipulation capability of metasurfaces. Here, we demonstrate an electrically tunable multifunctional metasurface in the visible range by integrating birefringent liquid crystals (LCs) with all-dielectric metasurfaces based on a novel packaging scheme. By combining the helicity-dependent geometric phase of the metasurface and the polarization control ability of LC molecules, continuous intensity tuning and switching of two helicity channels are realized. Electrically tunable single-channel switchable metaholograms, multicolor multiplexed metaholograms, and dynamic varifocal metalenses are designed to verify the concept. The exploration of polarization control in dynamic tuning can pave the way for dynamic metasurface devices in various applications, such as space light modulators, light detection and ranging systems, and holographic displays.

**Introduction**

Metasurfaces, two-dimensional metamaterials engineered by subwavelength metaunit arrays, exhibit an unprecedented capability to manipulate the amplitude, phase, frequency, and polarization of the electromagnetic wave. Various metasurface-based optics, such as beam shapers [1-2], polarization converters [3], metaholograms [4-6], metalenses [7-8], vectorial beam generators [9-10], and quantum metasurfaces [11-12], have been recently demonstrated. Among them, all-dielectric metasurfaces have attracted immense interest because of their high efficiency and complementary metal oxide semiconductor-compatible fabrication process, which holds great potential for their practical applications. Nevertheless, once fabricated, metasurfaces remain static, whereas tunability is desired in the near-infrared and visible range to realize dynamic holographic displays, optical communications, beam steering in light detection and ranging (LiDAR) systems, transparent displays, and automatic zoom lenses.

To date, tunable metasurfaces have been extensively studied. Particularly, programmable digital metasurfaces have been successfully demonstrated in the microwave range with active elements [13-15]; however, expanding this concept to optical metasurfaces is a challenge because of their small pixel size and unsuitable modulating mechanisms. Therefore, many other tuning strategies have been explored, such as the use of phase-change materials [16-17], free-carrier effects [18-20], mechanical actuation [21-22], chemical reaction [23-25], and manipulation of the surrounding environment [23-25]. Among them, liquid crystals (LCs) can provide a tangible solution for the dynamic control of local birefringence by exerting external electric fields to achieve tunable metasurfaces

[29-32]. However, the present research is mostly limited to the modulation of the overall resonance frequency based on the manipulation of the complex refractive index of the surroundings by directly wrapping nanofins in LCs. The polarization manipulation ability of tunable birefringent LCs and anisotropic metaunits has not been fully combined. Additionally, this integration strategy relying on direct wrapping faces the following challenges: First, the high refractive index of LCs reduces the refractive index contrast between the dielectric material and the environment, decreasing the efficiencies and performances of the propagating- or geometric phase metasurfaces [33]. Second, the small refractive index variation in LCs limits the dynamic tuning range[29]. Third, dense nanofins heavily affect wettability, preventing the infiltration of viscoelastic LCs into metasurfaces [33].

In this work, we demonstrate an electrically tunable multifunctional $TiO_2$ metasurface (ETMM) in the visible range by integrating birefringent LCs with geometric phase metasurfaces based on a novel packaging strategy. By combining the helicity dependence of the geometric phase with the polarization manipulation function realized by the precisely designed phase retardation of LCs, the switching of two orthogonal polarization channels is electrically controlled by tailoring their intensities, enabling dynamic devices with monochrome holographic switch, dual-color holographic switch, and varifocal metalens. Moreover, our integrated scheme offers a simpler device preparation process, better LC performance, and higher device efficiency in comparison with the direct wrapping scheme. Additionally, the mechanism of the proposed device is based on the interaction of a single metaunit with LC molecules so that the electrodes can be used for tuning individual metaunits to enable programmable digital metasurfaces in the optical range.

**Results**

**Design of the Electrically Tunable Multifunctional Metasurfaces**

Figure 1 shows the schematic of the proposed ETMM. The main structure of the device is vertically stacked by an all-dielectric metasurface and nematic LCs (NLCs). The underlying metasurface is composed of geometric phase-based metaunits wrapped in materials with refractive indices lower than that of the LC, leading to enhanced refractive index contrast and device efficiency compared with those obtained by the direct wrapping of LCs. The geometric phase, also known as the Pancharatnam–Berry phase, is related to the metaunit orientation only, and its sign can be flipped by switching the helicity of the incident circular polarization (CP) of light, enabling helicity-multiplexed functions. A flat layer of $TiO_2$ at the top of the metasurface ensures the homogeneity of the metasurface and prevents metaunit orientation from affecting the pre-orientation angle of the LC, making it possible to add an alignment layer underneath the LC. This results in the uniform LC orientation that is also hard to achieve with the direct wrapping scheme of LCs. The upper LC layer is composed of two alignment layers with LC molecules in between. By precisely constructing the thickness of the LC layer, a variable wave plate (VWP) with different phase retardations can be realized by controlling the orientations of LC molecules through electric regulation. The dynamically LC-based VWP continuously manipulates the CP generated by the geometric phase metasurface, so the continuous intensity tuning of individual CP channel or the switch between two orthogonal CP channels can be achieved by filtering the linear polarizer. Additionally, the broadband feature of the geometric phase enables the construction of multiwavelength functionalities. Consequently, Figure 1 demonstrates monochrome metahologram switch, switchable multicolor hologram, and varifocal metalens.

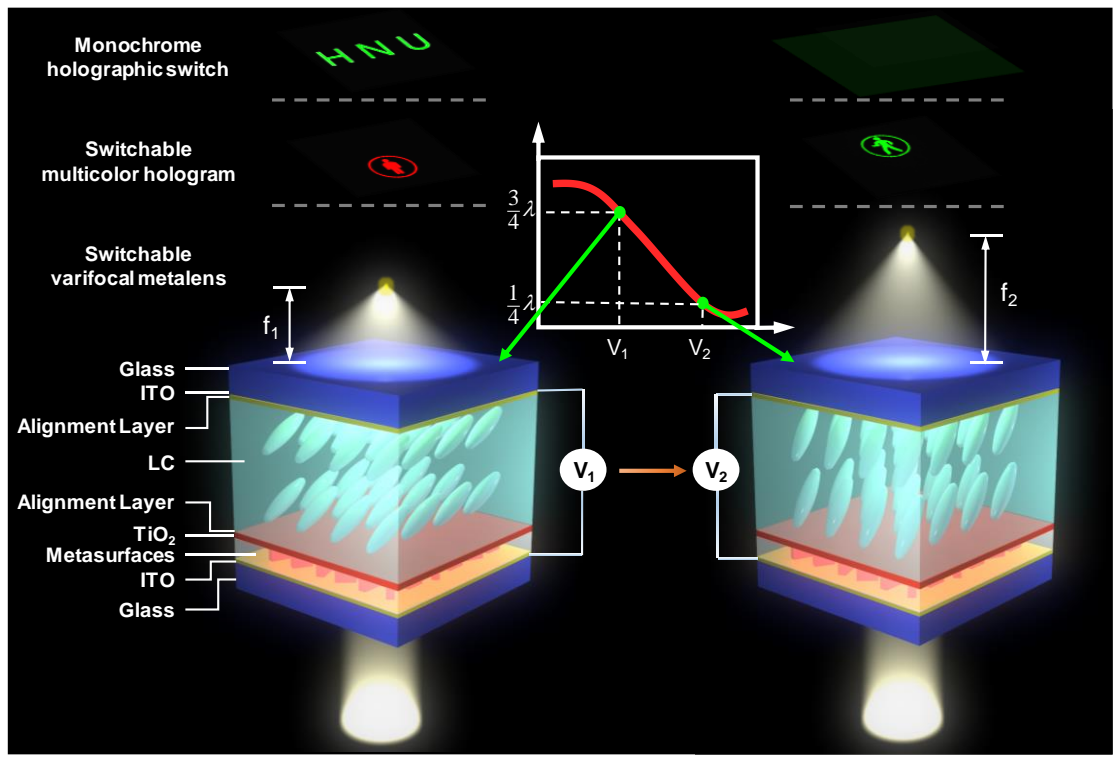

*Figure 1. Schematic of the electrically tunable multifunctional TiO$_2$ metasurface (ETMM). The ETMM is integrated by birefringent nematic liquid crystals (NLCs) with all-dielectric metasurfaces. The underlying metasurface is composed of geometric phase-based TiO$_2$ nanofins wrapped in PMMA and a flat layer of TiO$_2$ above the metasurface. The sample is encapsulated in a thin LC cell, which is covered by two alignment layers for the pre-orientation of LC molecules. By applying different voltages to the ITO at both sides, the orientation of LC molecules is adjusted to realize a variable wave plate with different phase retardations. Monochrome holographic switch, multicolor holographic switch, and varifocal metalens are*

*experimentally demonstrated.*

Figure 2a shows the detailed working principle of the ETMM. The metasurface is constructed by TiO$_2$ nanofins of the same size on a quartz substrate. TiO$_2$ is a dielectric material with a high refractive index in the visible range and can support large refractive index contrast to provide high efficiency at low nanofin height. The nanofins with in-plane structural parameters ($W$, $L$, $\theta$) and height ($H$) are arranged in square units with a period ($p$), forming metaunits. For the incident CP of light, by tuning the orientation angle of the nanofin at a coordinate $(x, y)$, the local phase change of the metasurface $\varphi(x, y)$ is modulated by $\varphi(x, y) = \pm 2\theta(x, y)$, known as the geometric phase carried by a transmitted beam with opposite helicity (see Supporting Information, Section 1). "+" and "−" represent the sign of the carrying phase for the incident right CP (RCP) and left CP (LCP), respectively. Therefore, if the helicity of the CP channel is dynamically changed, phase profile reversion can be achieved, resulting in the switching function.

The LC layer offers dynamic polarization conversion by enabling polarization channel selection with a linear polarized analyzer. The NLC is used with a refractive index of the ordinary axis $n_o = 1.529$ and extraordinary axis $n_e = 1.713$. The LC molecules are first integrated on the metasurface with the in-plane azimuthal angle α, depending on the alignment layer and the out-of-plane orientation angle $\theta_{LC} = 0°$, as shown in Figure 2a. If the electric field is applied between two electrodes sandwiching the LC, the anisotropy axis of the LC is switched from in-plane to out-of-plane. Consequently, the incident light with a polarization direction parallel to the in-plane projection of the LC anisotropy axis impinging normally onto the device experiences variable refractive index (i.e., equivalent refractive index $n$), which can be calculated by

$$\frac{1}{n^2} = \frac{\cos^2\theta_{LC}}{n_e^2} + \frac{\sin^2\theta_{LC}}{n_o^2} \qquad (1)$$

Because of the variable anisotropic refractive index, we can construct a wave plate with a variable phase retardation. For a given wavelength, the maximum amount of phase retardation is determined by the thickness of the LC layer, $d$, which can be written as $\Gamma = \frac{2\pi d(n-n_o)}{\lambda}$, where $\Gamma$ is the phase retardation and $\lambda$ represents the wavelength of the incident light. Note that the fast axis of the VWP is along the in-plane orientation angle, $\alpha$.

With this LC-based VWP, the CP emitted from the metasurface can be converted to variable elliptically polarized light, including the special cases of linearly polarized light with orthogonal polarization directions of $\alpha \pm 45°$ (Supporting Information, Section 2). Hence, if a linear polarizer with a direction of $\alpha + 45°$ or $\alpha - 45°$ is applied, the single CP channel can be switched. Moreover, since the CP with orthogonal helicity can be converted to orthogonal linearly polarized light, it is possible to switch the LCP and RCP channels with helicity-multiplexed metasurface and realize a dynamic multifunctional device. Figure 2b shows the theoretical results of the relationship between the phase retardation, intensity of output light, and out-of-plane orientation angle of LC molecules, $\theta_{LC}$, for the full-wave VWP (see Supporting Information, Section 2). We assumed that the incident light is LCP, the fast axis direction of the VWP is along the $x$ axis, and the linear polarizer direction at the exit end is 45°. As $\theta_{LC}$ changes from 0° to 90°, the phase retardation decreases, and the converted RCP light from metasurface is gradually transformed by the interaction with the VWP. Meanwhile, the intensity of the output light changes because of the filtering of the linear polarizer. When the phase retardation is $\frac{1}{2}\pi$ or $\frac{3}{2}\pi$, the two special orthogonal polarizations combined with the linear polarizer can achieve intensity switching. In the actual LC device, because of the anchoring effect of the alignment layer, the orientations of some LC molecules, especially those with alignments close to that of the film, cannot be changed even if a voltage is applied. This makes the distribution of LC molecules nonuniform, causing residual retardation during operation, which is ignored in our devices. In practical applications, the compensator can be used to meet the true zero-delay requirements in sensitive applications.

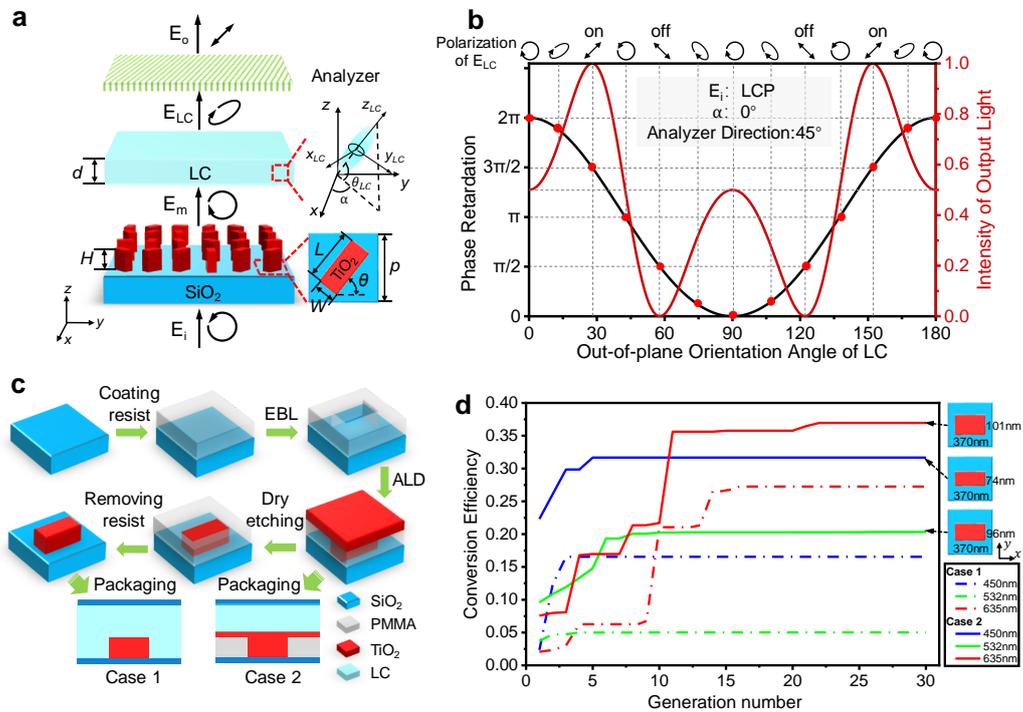

*Figure 2. Design and fabrication of the ETMM.* (a) The basic working principle of ETMM. The incident LCP modulated by the geometric phase-based metasurface carries the opposite

*helicity. The LC layer here is used for dynamic polarization conversion, enabling polarization channel selection with a linear polarized analyzer. (b) Theoretical results of the relationship between the phase retardation, intensity of emitting light, and out-of-plane orientation angle of LCs. It is assumed that the incident light is LCP; the in-plane orientation angle of LC molecules, α, is aligned to 0°; and the linearly polarized analyzer direction is 45°. (c) Fabrication of integrated metasurfaces via two different packaging strategies. (d) Conversion efficiency evolution by the particle swarm optimization (PSO) algorithm at λ = 450, 532, and 635 nm.*

As presented in Figure 2c, the device is mainly prepared by electron-beam lithography (EBL), atomic layer deposition (ALD), and LC packaging. The details of the fabrication processes and parameters can be found in the Methods section. To achieve the maximum polarization conversion efficiency (PCE) defined as the ratio between the optical power of the transmitted light with opposite helicity and the incident optical power, the nanofins should ideally act as half-wave plates (see Supporting Information, Section 1). This is achieved by birefringence, depending on the asymmetric cross section of nanofins with appropriately designed width, length, and height in a specific medium. To study the PCE under a given nanofin height, we explored two fabrication schemes for combining LCs with metasurfaces, as shown in Figure 2c. In Case 1, the metasurface is fabricated using the processes mentioned above, followed by LC packaging to fill the gaps of the nanofins. In Case 2, the LC is directly packaged after the ALD process, and the nanofins are surrounded by the resist. Compared with Case 1, Case 2 is much simpler. And because the upper $TiO_2$ layer by ALD can act as a flatten layer, an alignment layer can be added on the side close to the metasurface, which has great benefit for the uniformity of the LCs.

To compare the maximum PCE that can be achieved between the two schemes, we simulated the two cases by the finite-difference time-domain (FDTD) method and the built-in particle swarm optimization (PSO) algorithm to optimize the PCE of the final integrated device unit (see "Methods"). The period and the height of metaunits are set as 400 and 500 nm, respectively. Based on the entire search optimization, as the number of iterations increases, the efficiency eventually approaches the final optimal value. By defining the objective function and constraint conditions, the geometric parameters of each unit in the R, G, and B channels can be obtained. Figure 2d shows the PCE of nanofins for the two cases at 450, 532, and 635 nm. The optimized PCE of Case 2 is about twice higher than that of Case 1 at all wavelengths because the PMMA resist has a lower refractive index (1.48) than that of LC, resulting in a large refractive index contrast between PMMA and $TiO_2$, which makes the nanofin work closely as a half-wave plate under the same structural height. Therefore, considering the simplicity of processing and device stability and efficiency, Case 2 is chosen for device preparation. According to the optimized results, the lengths of the nanofins used are 370 nm, and their widths are 101, 96, and 74 nm for R, G, and B channels, and the corresponding maximum PCEs are 37%, 20%, and 32%, respectively. Note that the PCE can be further improved by increasing the heights of the nanofins.

**Concept Demonstration**

To prove the concept, we first prepared a device for a single-channel hologram switch. To obtain the orientations of the nanofins determining the phase profiles for the hologram image, the Gerchberg−Saxton (GS) algorithm was applied in a computer-generated hologram. The letter "HNU" was designed to encode in a 200 μm × 200 μm metasurface at a wavelength of 532 nm and was observed in the exiting LCP channel converted from RCP. The thickness of the LC is set as 2.9 μm for full-wave VWP at 532 nm. Figure 3a shows the picture of the device comprising the LC-integrated metasurface and electrode leads. The optical image of the metasurface area in the transmission mode shows excellent transmittance and uniformity after LC packaging, ensuring the remarkable performance of the device. Different layers can be distinguished in the cross-sectional view of the scanning electron microscope (SEM) images of the metasurface in Figure 3b. Among them, PMMA shrinks because of high-energy electron-beam irradiation, resulting in holes. To better observe the morphology of the metasurface, we fabricated the metasurface through the complete process in Figure 2c, and its top-view and oblique-view SEM images are shown in Figure 3c and d. The optical experimental setup for the holographic reconstruction is shown in Figure 3e. A linear polarizer (P1) and a quarter-wave plate act as an RCP polarizer. The reconstructed image is formed by collecting transmitted beams using an objective lens, filtering by a polarization analyzer (P2) with 45° direction, and finally, capturing by a CCD camera. Figure 3f shows the switching performance of holography under different voltages. The curve in the figure is calibrated by the optical power meter and the front and rear polarizers (see Methods). By applying the calibrated voltage to the device, the two switching states at phase retardations of $\frac{1}{2}\pi$ and $\frac{3}{2}\pi$ can be observed. The intensity change under different voltages satisfies the theoretical trend in Figure 2b. The continuous voltage tuning results in the continuous holographic intensity change (see Supplementary Video 1), which verifies the design principle. The background light in the off state is the unconverted RCP, which can be weakened by further improving the PCE.

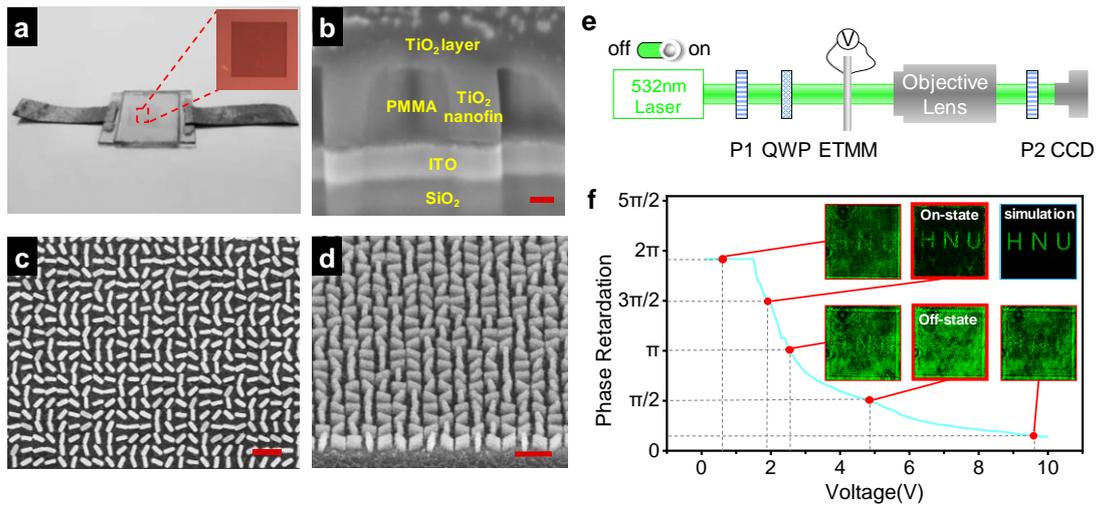

*Figure 3. Single-channel hologram switch achieved by the ETMM. (a) Photograph of the ETMM and LC cell with a uniform LC layer thickness (upper right corner). (b) Cross-sectional view of the scanning electron microscope (SEM) images of the metasurface. The scale bar is 100 nm. (c and d) Top-view and oblique-view SEM images of the metasurface processed through the complete process shown in Figure 2c. The scale bar is 1 μm. (e) Optical setup for holography reconstruction. (f) Switching performance of holography under different voltages. The thickness of the LC is 2.9 μm.*

**Electrically Tunable Helicity-Multiplexed Metaholograms**

Since the designed devices can realize the switch of two orthogonal helicity channels, the multiplexed switchable functions can be realized if we encode the information into both helicity channels. Interleaved geometric phase metasurfaces [34], the vector decomposition of the geometric phase with noninterleaved metasurfaces [35], and a combination of propagation and geometric phases [36] have been applied to achieve helicity-multiplexed monochrome metaholograms. Here we adopted the vector decomposition scheme with noninterleaved metasurfaces composed of single-size nanofins to achieve the helicity-multiplexed switchable multicolor holograms. Figure 4a shows the vector decomposition principle of the geometric phase by a single metaunit for two wavelengths. Based on the broadband property of the geometric phase and the dispersion relationship of propagation, two phase profiles are retrieved by the GS algorithm: One phase profile, $\varphi_{LCP}^{\lambda_1}(x,y)$, is designed to reconstruct the holographic image ("stop signal" of traffic lights as an example) at a given z plane for wavelength $\lambda_1$ under LCP; another phase profile, $\varphi_{RCP}^{\lambda_2}(x,y)$, is designed to reconstruct the holographic image ("pass signal" of traffic lights as an example) at the same plane for wavelength $\lambda_2$ under RCP. Then, the geometric phase, $\varphi(x,y)$, of the metasurfaces can be encoded by a vector decomposition: $Ae^{i\varphi(x,y)} = e^{i\varphi_{RCP}^{\lambda_2}(x,y)} + e^{-i\varphi_{LCP}^{\lambda_1}(x,y)}$. The final phase profile, $\varphi(x,y)$, is encoded into the metasurface by nanofins with an orientation angle of $\varphi(x,y)/2$ for RCP. Consequently, when the RCP beams of $\lambda_1$ and $\lambda_2$ are incident on a metasurface, a real image and a virtual image exist for each beam in exiting LCP channel that is symmetric about the position of the metasurface as shown on the right of Figure 4a. Because of the dispersion, only the image of the designed wavelength is on the given plane. When a combination of input/output polarization is switched, the locations of the real image and the virtual image are swapped. Consequently, the two independent real images ("stop signal" and

"pass signal") for different colors on the same position are switchable, depending on the helicity channel. Since the LCs do not function as a broadband wave plate, the switch voltages of different wavelengths are different and are separately calibrated. When the device is working in a certain wavelength, the on-voltage of that wavelength is applied. At this time, the other wavelength channel is not completely turned on, which is beneficial to eliminate the crosstalk (see Supplementary Information, Section 3). Figure 4b shows the optical experimental setup for characterizing the multicolor switchable holographic metasurfaces. A dichroscope (D) is used to combine the lasers in one light path. A linearly polarized beam containing equal components of LCP and RCP is incident on the sample. First, we designed devices composed of 200 μm × 200 μm metasurfaces for the special case in which the two wavelengths are the same, resulting in a monochrome switchable hologram (Figure 4c). Independent holographic images of "smiling face" and "crying face," "Chinese Fu" and "rat" are encoded into two devices. The designed hologram reconstruction plane is 200 μm away from the sample surface. The results of switching with a wave plate before packaging and voltage-tunable switching after packaging are compared in Figure 4c. The switch results are in good agreement with the simulation verifying the helicity-multiplexed switchable functions (see Supplementary Video 2 for continuous switch performance). Furthermore, we demonstrated two devices for a multicolor hologram switch in Figure 4d. The device size and holographic reconstruction distance remain the same. The two traffic light signs and the two leaf patterns are respectively encoded into two devices at 635 nm (red) and 532 nm (green) wavelengths. We can see that only one of the images of designed wavelengths can be observed, whereas the image of the other wavelength is almost invisible since it is not on the reconstruction plane, and the on-state voltage is inconsistent. This can effectively avoid crosstalk to achieve multicolor dynamic holography, resulting in new opportunities for dynamic holographic displays.

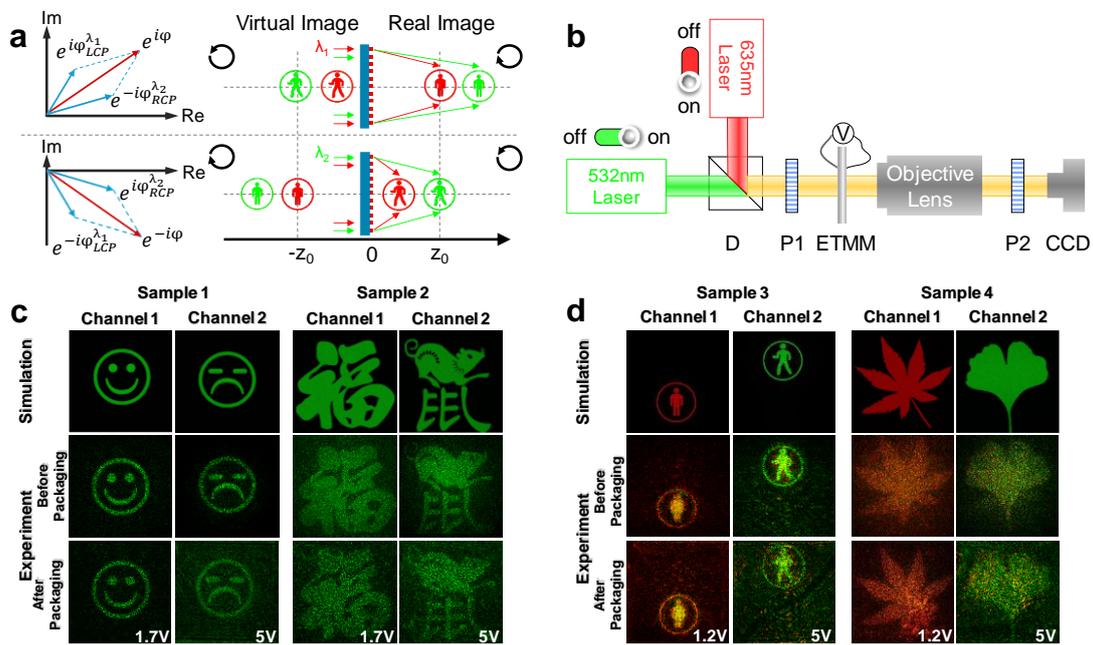

*Figure 4. Demonstration of electrically tunable helicity-multiplexed metaholograms. (a) Vector decomposition of the geometric phase by a single nanofin. Under the illumination of LCP (RCP), the phase profile reconstructs the holographic image "stop signal" ("pass signal") at the observation plane z0 under a 635 nm (532 nm) beam. (b) Optical setup for characterizing multicolor switchable holographic metasurfaces. (c and d) Simulated and experimental results for the single-color hologram switch and the multicolor hologram switch, respectively. The simulations and the results of wave plate switching before packaging and voltage-tunable switching after packaging are shown in the top, middle, and bottom rows, respectively.*

**Electrically-Tunable Varifocal Metalenses**

Using the same vector decomposition scheme described above, we also demonstrate an electrically tunable varifocal metalens. Two independent spherical phase profiles can be encoded into a single geometric phase metasurface and switched with different helicity channels for two focal points. If two axial focal points are designed, the electrically tunable zoom imaging can be realized. We prepared a device with metalens (200 μm diameter) at 532 nm wavelength for switching the focal points at 200 and 400 μm, resulting in NAs of 0.45 and 0.24, respectively. The focal point spread functions at three wavelengths along the propagation direction (z-axis) were measured with a 1 μm resolution (Figure 5a). It can be seen that the switching performance at each wavelength shows a good broadband property, whereas the focal points of different wavelengths shift similar to a Fresnel lens because of the dispersion. Figure 5b shows the normalized intensity profiles of the two focal points for each wavelength. The metalens is diffraction-limited, and the focal point becomes larger as the focal length increases or the wavelength increases. To demonstrate the use of the electrically tunable varifocal metalens for practical zoom imaging, we characterized the imaging resolution using the 1951 United States Air Force resolution chart as the target object (see the measurement configuration in Figure S2). In the test, we fixed the object distance to 1200 μm and adjusted the voltage to observe the image formed by two focal points at image distances of 240 and 600 μm for the zoom factor of 1/5 and 1/2 (the relationship is $\frac{1}{v} + \frac{1}{u} = \frac{1}{f}$, where $v, u,$ and $f$ are the image distance, object distance, and focal length, respectively) at 532 nm. Table S1 shows the object and image distance parameters for the other two wavelengths. Figure 5c provides the results of the imaging. It can be seen that the device can reasonably perform electronically tunable zoom imaging in the visible range. The device was also characterized by measuring the focusing efficiency of the focal points under different voltages. The focusing efficiency is defined as the ratio of the focal point power to the transmitted power through an aperture with a diameter that is the same as that of the metalens. The measured focusing efficiencies in Figure 5d illustrate two focal points switching with voltage variation and the change trend corresponds to the theoretical curve in Figure 2b. The inset shows focal point profiles for different voltages.

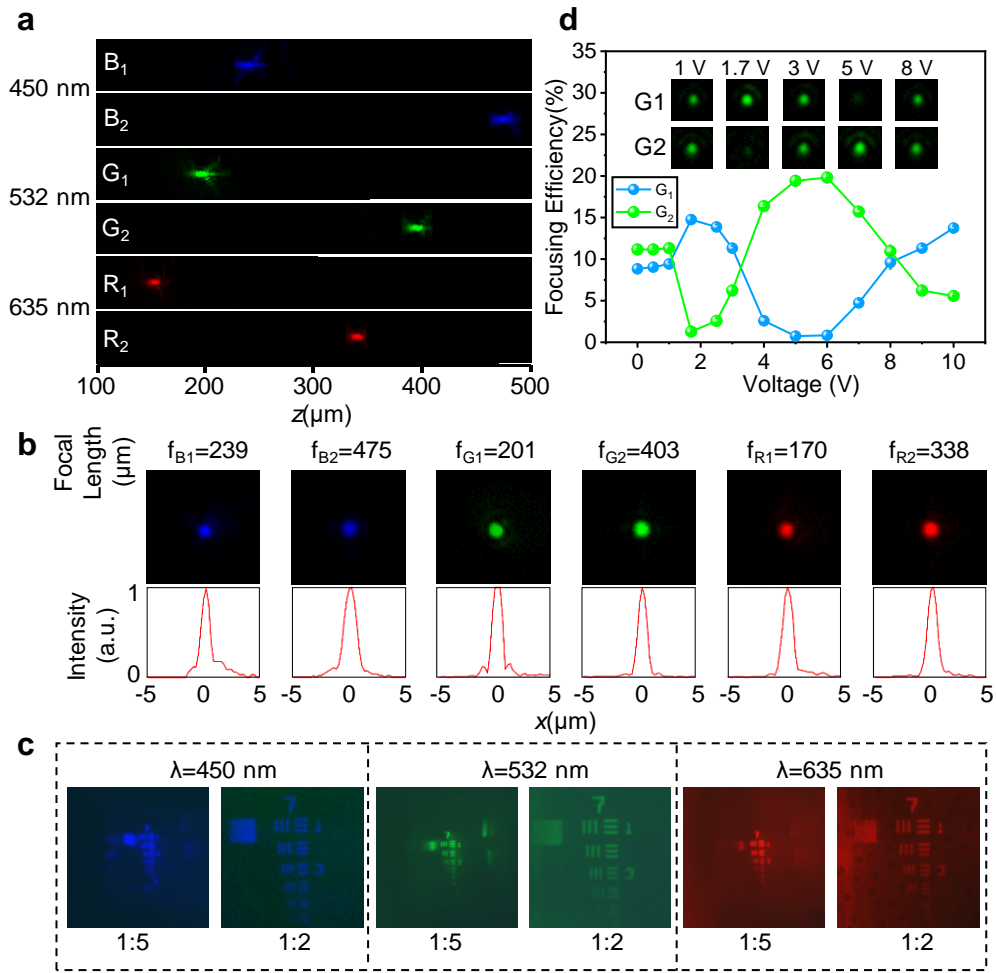

*Figure 5. Demonstration of electrically-tunable varifocal metalens.* (a) Experimental light intensity profiles in the x-z plane of the electrically tunable varifocal metalens at various wavelengths in the visible range (labeled to the left of the plots). The white dashed line indicates the position of the focal plane. (b) Focus image and normalized intensity profiles along the focal plane for various wavelengths. (c) Images of the 1951 United States Air Force resolution target formed by the electrically tunable varifocal metalens. (d) Relationship between the applied voltage and focusing efficiency of the electrically tunable varifocal metalens.

**Discussion**

The efficiency of our devices can be further improved by increasing the PCE with higher nanostructures and materials with higher refractive indices to transform the metaunits into a half-wave plate. Because the metasurface has an unprecedented capability of subwavelength-scale pixelated polarization control for polarization conversion and polarization-dependent multiplexed and even complex vector beams [37], many dynamic devices can be achieved if the polarization control capability of LCs is also utilized. Recently, pixelated electrical regulation has been applied to the phase-only dynamic metasurface [32], which can also be applied in our devices to achieve pixelated polarization control. Additionally, through the achromatic design of the geometric phase metalens [38-39], our zoom imaging can also be further improved to achromatic imaging. Since LCs can be electrically controlled and also respond to heat, force, light, and magnetic field, exciting possibilities emerge for their combination with metasurfaces.

**Conclusion**

In summary, we combined the polarization control ability of the geometric phase metasurface and the electronically controlled birefringence characteristics of LCs to achieve an electrically tunable multifunctional metasurface in the visible range. Utilizing the characteristics of the abrication method of metasurfaces, a new packaging scheme for metasurfaces and LCs is explored, which not only simplifies the processing but also improves the uniformity of the LCs. The device is optimized through the PSO algorithm to achieve maximum efficiency. Finally, electrically tunable single-channel switch metaholograms, multicolor multiplexed metaholograms, and dynamic varifocal metalenses were developed to verify the devices. Our device realizes dynamic metasurface devices in the visible range and can be further extended to dynamic applications, such as programmable spatial light modulators and LiDAR. More importantly, it paves the way for blending the polarization control capabilities of metasurfaces and LCs.

**Methods**

**Numerical simulation.** The polarization conversion efficiency (PCE) of the metasurface were simulated by the finite-difference time-domain (FDTD, Lumerical FDTD Solutions) method and the built-in Particle swarm optimization (PSO) algorithm. The period and height of meta-atoms were set as 400 nm and 500 nm. Considering the constraints of experimental conditions and period sizes, we set the minimum and maximum size constraints of the nanofins to be 50 nm and 370 nm, respectively. The nanofin arrangement mode for single period in the different cases was shown in Figure 2c. For the simulation, the refractive index of the $TiO_2$ was the measurement result by ellipsometer and the refractive indexes of PMMA and LC were set as 1.48 and 1.713, respectively. In order to get the optimal structure parameters and the maximum PCE, PSO algorithm with generation size of 10 and maximum generations of 30 was carried out. The termination condition of the iteration was defined as that the difference between the current optimal solution being calculated and the average value of the previous three generations is 0.

**Device preparation.** First, a 500 nm polymethyl methacrylate (PMMA) electron-beam resist layer was spin coated on transparent glass substrate with ITO film layer. Then, the resist was patterned by an EBL system (RAITH150 Two) with a 30-KV voltage and a beam current of 174 pA. Subsequently, the sample was developed in a mixed solution of methyl isobutyl ketone (MIBK) and isopropanol (IPA) (MIBK:IPA = 1:3) for 1 minute, and fixed in the IPA for 1 minute. After development, a 68nm $TiO_2$ was deposited by an ALD system filling in the voids in the resist from the sidewalls. In addition to filling these voids, this created an excess layer of $TiO_2$ of 68 nm on top of the nanofins. Therefore, ion beam etching (IBE) and reactive ion etching (RIE) are required to remove the top layer $TiO_2$ and resist at the end. In the fabrication of device, the LC was directly packaged after the ALD process (Case 2 in the Figure 2c). For the LC packaging process, the prepared substrate after ALD process and another pure ITO glass substrate are used as the two sides of the LC cell. Both the two substrates are spin-coated with azo-dye photo-alignment material SD1 (from Dainippon Ink and Chemicals Inc) and then assembled as a cell with controlled cell gap 2.9 μm by spacers. The prepared LC cell is then exposed to polarized blue light (450 nm) to generate an initial alignment direction. The commonly used LC material 5CB is used to fill into the LC cell, thus the device in Case 2 in the Figure 2c is fabricated.

**Optical characterization.** Details of the optical experimental setup for characterizing the switchable metalens is shown in Figure S3. The curve in the Figure 3f was calibrated by the optical power meter and the front and rear polarizers at 532 nm. Three laser diodes emitting at 450 nm, 532 nm and 635 nm were utilized for the calibration of R, G and B channels, respectively.

**Acknowledgments**

The authors thank Mrs. Rong Wang for sectional SEM image preparation. The authors acknowledge the financial support by the National Natural Science Foundation of China under contract no. 52005175, 51722503, 5162100, and Natural Science Foundation of Hunan Province of China under contract number 2020JJ5059..

**Author contributions:**

Y.H. and H.D. proposed the idea. Y.H., X.O. conceived and carried out the design and simulation. X.O., J.L., X.L. (Xuhao Luo) and X.L. (Xin Li) prepared the metasurface samples. T.Z. and F.F. carried out the LCs packaging. Y.H., X.O., J.Z., T.Z., L.L. conceived and performed the measurements. Y.H., X.O., T.Z., L.L., F.F. and H.D. analyzed the results and co-wrote the manuscript. All the authors discussed the results and commented on the manuscript.

**Competing interests:** The authors declare that they have no competing interests.